\batchmode
\makeatletter
\def\input@path{{/Users/ananyabalakrishna/Desktop/}}
\makeatother
\documentclass[10pt,twocolumn]{article}
\usepackage[latin9]{inputenc}
\usepackage[letterpaper]{geometry}
\usepackage{color}
\usepackage{float}
\usepackage{amsmath}
\usepackage{amssymb}
\usepackage{graphicx}
\usepackage{setspace}
\usepackage[numbers]{natbib}
\usepackage[unicode=true,pdfusetitle,
 bookmarks=true,bookmarksnumbered=false,bookmarksopen=false,
 breaklinks=false,pdfborder={0 0 1},backref=false,colorlinks=false]
 {hyperref}

\makeatletter
\usepackage{ragged2e}
\usepackage{booktabs}
\usepackage{authblk}
\usepackage{setspace}

\usepackage[parfill]{parskip}
\usepackage{graphicx}
\usepackage{cite}
\usepackage{notoccite}

\title{Combining phase field crystal methods with a Cahn-Hilliard model for binary alloys}
\author{Ananya Renuka Balakrishna, W Craig Carter}
\affil{Department of Materials Science and Engineering, Massachusetts Institute of Technology, Cambridge, MA02139, USA}

\date{}                    

\@ifundefined{showcaptionsetup}{}{%
 \PassOptionsToPackage{caption=false}{subfig}}
\usepackage{subfig}
\makeatother

\begin{document}
\twocolumn[   
\begin{@twocolumnfalse}  

\maketitle 
\begin{abstract}
\begin{singlespace}
\textcolor{black}{During phase transitions certain properties of a
material change, such as composition field and lattice-symmetry distortions.
These changes are typically coupled, and affect the microstructures
that form in materials. Here, we propose a 2D theoretical framework
that couples a Cahn-Hilliard (CH) model describing the composition
field of a material system, with a phase field crystal (PFC) model
describing its underlying microscopic configurations. We couple the
two continuum models via coordinate transformation coefficients. We
introduce the transformation coefficients in the PFC method, to describe
affine lattice deformations. These transformation coefficients are
modeled as functions of the composition field. Using this coupled
approach, we explore the effects of coarse-grained lattice symmetry
and distortions on a phase transition process. In this paper, we demonstrate
the working of the CH-PFC model through three representative examples:
First, we describe base cases with hexagonal and square lattice symmetries
for two composition fields. Next, we illustrate how the CH-PFC method
interpolates lattice symmetry across a diffuse composition phase boundary.
Finally, we compute a Cahn-Hilliard type of diffusion and model the
accompanying changes to lattice symmetry during a phase transition
process.}
\end{singlespace}

\medskip{}
\end{abstract}
\end{@twocolumnfalse} ]
\begin{singlespace}

\section*{{\small{}Introduction}}
\end{singlespace}

{\small{}Phase transitions in materials are typically accompanied
by structur}\textcolor{black}{\small{}al changes to the lattice symmetry
\citep{LinesM.E.1977,James1986,Meethong2008}. These changes include
individual lattice distortions \citep{LinesM.E.1977,Meethong2008},
grain rotations \citep{James1986} and lattice defect formations \citep{LevanyukA.P.&Sigov1988}.
All of these influence the microstructures that form in a material
\citep{LevanyukA.P.&Sigov1988,SongY.ChenX.DabadeV.ShieldT.W.&James2013}.
In this paper, we introduce a m}{\small{}odeling approach that couples
the lattice symmetry with the composition field. Two continuum methods,
namely a Cahn-Hilliard model and a phase field crystal model, are
coupled and solved to describe a phase transition process.}{\small \par}

{\small{}The lattice symmetry and grain orientations in microstructures
are known to affect physical properties of materials. For example,
lithium diffusion in a battery electrode induces lattice deformation
}\textcolor{black}{\small{}\citep{Nie2014,YuanY.NieA.OdegardG.M.XuR.ZhouD.SanthanagopalanS.HeK.Asayesh-ArdakaniH.MengD.D.KlieR.F.andJohnson,Tang2006a,Warren2003a},
which affects lithium ion kinetics \citep{Nie2014} and causes anisotropic
expansion of electrodes \citep{YuanY.NieA.OdegardG.M.XuR.ZhouD.SanthanagopalanS.HeK.Asayesh-ArdakaniH.MengD.D.KlieR.F.andJohnson}.
Likewise, in a paraelectric to ferroelectric phase transitions, lattices
transform from centrosymmetric to other point groups lacking an inversion
centre. This transformation introduces stress-free spontaneous strains
in the ferroelectric system \citep{LinesM.E.1977}. At present, theoretical
models like phase field methods describe complex microstructures in
electrode/ferroelectric systems as a function of the composition field
(lithium-ion concentration, temperature or polarization) \citep{Tang2009,Cogswell2012a,Chen2002,Balakrishna2016}.
The Kobayashi-Warren-Carter phase field model \citep{Warren2003a}
further accounts for crystallographic misorientation at grain boundaries
during a phase transition process. While these modeling approaches
provide insights on the position of phase and grain boundaries, they
only account for grain orientations as an empirical parameter \citep{Warren2003a}.
The current phase field approaches do not allow for lattices to distort
independently. Consequently, the local strain fields arising from
individual lattice distortions and the presence of defects in a material
system are not explored.}{\small \par}

\textcolor{black}{\small{}Alternatively, a phase field crystal (PFC)
method proposed by Elder and Grant \citep{Elder2002a,Elder2004a}
describes atomistic details of material systems with periodic solutions.
This modeling technique describes coarse-grained symmetry of a periodic
system \citep{Emmerich2012b,Provatas2007}, and is computed at faster
time scales than the molecular dynamics simulations \citep{Tupper2008}.
The PFC model has been applied to explore lattice defects in graphene
\citep{Seymour2016b} and nucleation problems in colloidal systems
\citep{GranasyL.TegzeG.TothG.I.andPusztai2011}. Binary alloy models
\citep{ElderK.R.ProvatasN.BerryJ.StefanovicP.andGrant2007a,ElderK.R.HuangZ.F.andProvatas2010,KundinJ.ChoudharyM.A.andEmmerich2014,AlsterE.ElderK.R.HoytJ.J.andVoorhees2017},
an extension to the PFC formalism, was demonstrated to describe solidification
\citep{KundinJ.ChoudharyM.A.andEmmerich2014,HeinonenV.AchimC.V.ElderK.R.BuyukdagliS.andAla-Nissila2014},
crystallization \citep{ElderK.R.ProvatasN.BerryJ.StefanovicP.andGrant2007,ZhangW.andMi2016}
and phase segregation processes \citep{TaoY.ZhengC.JingZ.Wei-PingD.andLin2012,GreenwoodM.SinclairC.andMilitzer2012}.
The PFC approach is a useful tool for multiscale modeling to describe
the lattice symmetry of a material system.}{\small \par}

\textcolor{black}{\small{}In the current work, we combine the phase
field crystal methods with a Cahn-Hilliard model in a 2D theoretical
framework to model a phase transition process. The modeling approach
couples two field parameters of a model system, namely the composition
field and the coarse-grained lattice-symmetry distortions. The Cahn-Hilliard
equation describes microstructures with a composition order-parameter
field. The phase field crystal equation models a coarse-grained representation
of lattice symmetry with peak density field as its order parameter.
In the PFC equation, we introduce coordinate transformation coefficients
to relate lattice symmetries in 2D point groups via affine transformations.
These transformation coefficients are coupled with the composition
field and influence the underlying lattice symmetry of a material
system. As the composition field evolves following the Cahn-Hilliard
equation, the transformation coefficients are updated in the PFC model.
The PFC model computes the equilibrium lattice arrangements of the
material system during composition evolution. Here, an assumption
is that the dynamics of the PFC model is fast relative to the composition
field dynamics. Using this coupled approach, we model the structural
evolution of lattice distortions and defects during a phase transition
process.}{\small \par}

\textcolor{black}{\small{}In this paper, we investigate the nature
of the coupled CH-PFC methods by modeling three representative examples.
First, we consider base cases to understand how transformation coefficients
affect the coarse-grained lattice symmetries of the material system.
Here, we stabilize hexagonal and square symmetries as representative
lattice structures corresponding to two composition-field values.
Second, we extend these base cases to investigate how the model interpolates
the peak density field across a diffuse composition phase boundary.
We model a representative binary alloy with hexagonal and square symmetry
phases and explore the lattice distortions across diffuse interfaces.
Finally, we model the composition field in the binary alloy to follow
a Cahn-Hilliard type of diffusion and study the accompanying equilibrium
lattice arrangements described by the peak density field. The simulations
show lattice distortions at coherent interfaces and demonstrate structural
evolution of lattice arrangements during a phase transition.}{\small \par}

\section*{\textcolor{black}{\small{}Coupled Cahn-Hilliard \textendash{} phase-field
crystal model}}

\textcolor{black}{\small{}The aim is to couple the Cahn-Hilliard (CH)
and the phase field crystal (PFC) methods, to explore structural changes
to lattice symmetry during diffusion induced phase transition. In
this section, we first introduce the two continuum models and explain
how these methods are coupled in a 2D theoretical framework. Next,
we describe the evolution of the two order parameters, namely the
composition field and the peak density field. Finally, we discuss
the numerical procedure followed to compute the coupled CH-PFC methods.}{\small \par}

\textcolor{black}{\small{}The first model is a Cahn-Hilliard method
that describes the continuum composition field of a model system.
This method utilizes a double-well free-energy function}{\small{}
in terms of a composition field,}\textcolor{black}{\small{} $\overline{c}$,
which is its order parameter. The second model is a phase field crystal
(PFC) method that describes the coarse-grained lattice symmetry for
the model system, and statistically illustrates lattice orientation,
distortion and defect density. This approach describes a free energy
functional that is minimized by a spatially-periodic order parameter,
$\phi$. In the current work, we couple the two models by using the
composition field to influence the underlying lattice symmetry of
the model system. The composition is not coupled to the peak density
field $\phi$ via a homogeneous free energy, but rather as the coordinate
transformation coefficients of the composition-dependent Laplacian
$\nabla_{c}^{2}$, relative to a Cartesian basis. That is, each of
the 5 Bravais lattices in the 2-dimensional space are stabilized by
computing the Laplace operator in a transformed space on a coordinate
plane. The transformation coefficients which control lattice deformations
are described as functions of the composition field. These coefficients
are updated during the evolution of the composition field. }{\small \par}

\textcolor{black}{\small{}The total free energy functional for the
CH-PFC model is given by:}{\small \par}

\textcolor{black}{\footnotesize{}
\begin{align}
F & =\int[g(\overline{c})+\kappa|\nabla\overline{c}|^{2}+f(\phi)+\frac{\phi}{2}G(\nabla_{c}^{2})\phi]d\vec{r}\nonumber \\
 & =\int[\frac{F_{0}}{(\overline{c}_{a}-\overline{c}_{b})^{4}}(\overline{c}-\overline{c}_{a})^{2}(\overline{c}-\overline{c}_{b})^{2}+\kappa|\nabla\overline{c}|^{2}\nonumber \\
 & \hspace{0.1cm}\hspace{0.1cm}\;+a\Delta T\frac{\phi^{2}}{2}+u\frac{\phi^{4}}{2}+\frac{\phi}{2}\hspace{0.1cm}(\lambda(q_{0}^{2}+\nabla_{c}^{2})^{2})\hspace{0.1cm}\phi]d\vec{r}\label{eq:1}
\end{align}
}{\footnotesize \par}

\textcolor{black}{\small{}Here, $g(\overline{c})$ and $f(\phi)$
describe the homogeneous energy contributions from the Cahn-Hilliard
and PFC equations respectively. The composition gradient-energy coefficient
is given by $\kappa$. The operator $G(\nabla_{c}^{2})$ controls
the coarse-grained lattice symmetry described by the particle density
field. This operator is modeled as a function of the composition field
and is discussed in detail later on in this section. The coefficients,
$F_{0}$ and $(\overline{c}_{a},\overline{c}_{b})$, correspond to
the energy barrier height and to the local equilibrium states of $g(\overline{c})$
respectively. The parameter $a\Delta T$ , controls the second-order
phase transition of the PFC model. In this paper, we model $a\Delta T$
as a constant to always describe a crystalline-solid state. The parameters
$\lambda,q_{0},u$, relate the PFC equation to the first-order peak
in an experimental structure factor. Further details on these coefficients
are explained in the work by Elder and Grant \citep{Elder2004a}. }{\small \par}

\textcolor{black}{\small{}Before proceeding with the model description,
we first normalize the free energy functional:}\\
\textcolor{black}{\footnotesize{}
\begin{align}
\mathcal{F} & =\frac{F}{F_{0}}\nonumber \\
 & =\int\{c^{2}(c-1)^{2}+|\nabla c|^{2}+\gamma[\frac{\psi}{2}(r+(1+\nabla_{c}^{2})^{2})\psi+\frac{\psi^{4}}{4}]\}d\vec{x}.\label{eq:2}
\end{align}
}{\footnotesize \par}

\textcolor{black}{\small{}The composition field, $\overline{c}$ is
normalized as $c=\frac{\overline{c}_{a}-\overline{c}}{\overline{c}_{a}-\overline{c}_{b}}$,
with local equilibrium states at $c=0$ and $c=1$. The dimensionless
peak density field, $\psi$ is given by $\psi=\phi\sqrt{\frac{u}{\lambda q_{0}^{4}}}$.
We set, $r=\frac{a\Delta T}{\lambda q_{0}^{4}}=-0.2$, as a constant
such that Eq. \ref{eq:2} always models a stable crystalline-solid
phase for the peak density field \citep{Elder2004a}. With $c=0$,
the peak density field in Eq. \ref{eq:2} describes a hexagonal symmetry
with a periodic spacing of }\textit{\textcolor{black}{\small{}$\frac{4\pi}{q_{0}\sqrt{3}}$}}\textcolor{black}{\small{}
at equilibrium. Note, $\frac{1}{q_{0}}$ is the length scale of the
PFC model and $\vec{x}=q_{0}\vec{r}$. The gradient energy coefficient
$\kappa=\frac{F_{0}}{(\overline{c}_{a}-\overline{c}_{b})^{2}}\left(\frac{16\pi}{q_{0}\sqrt{3}}\right)^{2}$,
is numerically calibrated such that the width of the diffuse composition
interface spans over $\sim4$ peaks described by the peak density
field, $\psi$. We introduce a constant, $\gamma=\frac{\lambda^{2}q_{0}^{5}}{uF_{0}}$
that relates the free energy normalizations of the Cahn-Hilliard and
the PFC model. For simulations in this paper, we set $\gamma=1$. }{\small \par}

\textcolor{black}{\small{}The composition-dependent Laplacian $\nabla_{c}^{2}$
in Eq. \ref{eq:1}, introduces the composition-lattice symmetry coupling.
Here, the composition terms enter the Laplacian via its coordinate
transformation coefficients. The Laplace operator is written in terms
of its second partial derivatives:}{\small \par}

\textcolor{black}{\small{}
\begin{equation}
\nabla_{c}^{2}=(\mathrm{A}_{1,1}^{2}+\mathrm{A}_{1,2}^{2})\frac{\partial}{\partial x^{2}}+\mathrm{A}_{2,2}^{2}\frac{\partial}{\partial y^{2}}+2\mathrm{A}_{1,2}\mathrm{A}_{2,2}\frac{\partial}{\partial x\partial y},\label{eq:3}
\end{equation}
}{\small \par}

\textcolor{black}{\small{}where $\mathrm{A}_{k,l}$ are the coordinate
transformation coefficients. These coefficients are described as functions
of the dimensionless composition field $c$ and correspond to the
elements of a $k\times l$ transformation matrix:
\begin{equation}
\mathbf{A(\mathit{c})}=\left[{\begin{array}{cc}
\alpha(c) & \frac{2\alpha(c)}{\sqrt{3}}\hspace{0.1cm}\mathrm{cos}[\theta(c)]-\frac{\alpha(c)}{\sqrt{3}}\\
0 & \frac{2\beta(c)}{\sqrt{3}}\hspace{0.1cm}\mathrm{sin}[\theta(c)]
\end{array}}\right].\label{eq:4}
\end{equation}
}{\small \par}

\textcolor{black}{\small{}The matrix $\mathbf{A(\mathit{c})}$, describes
affine lattice transformations using hexagonal symmetry as the reference
structure \citep{BallJ.M.andJames1987}, \citep{Bhattacharya2003}.
With $c=0$, the transformation matrix is an identity matrix and Eq.
\ref{eq:2} describes a hexagonal symmetry in 2D \citep{Elder2004a}.
In the current work, we choose the hexagonal and square symmetries
to represent phases with compositions $c=0$ and $c=1$ respectively.
These symmetries are chosen to illustrate exaggerated symmetry deformations
during phase transition. The transformation coefficients in Eq. \ref{eq:4},
$(X=\alpha,\beta,\theta)$, are modeled as linear functions of the
dimensionless composition field, $X(c)=X_{0}+c\Delta X$. We define
$X_{0}$ to be the transformation coefficients corresponding to the
hexagonal lattice $(\alpha_{0}=\beta_{0}=1,\theta_{0}=\frac{\pi}{3})$,
and $\Delta X$ is the deformation required to transform the lattice
with a hexagonal symmetry to a square symmetry $(\Delta\alpha=\Delta\beta=0,\Delta\theta=\frac{\pi}{6})$.
Note, in both the hexagonal and square lattice symmetries, the transformation
matrix encourages a periodic lattic- symmetry spacing of $\frac{4\pi}{q_{0}\sqrt{3}}$. }{\small \par}

\textcolor{black}{\small{}}
\begin{figure}[H]
\centering{}\textcolor{black}{\small{}}\subfloat[\label{fig:1a}]{\textcolor{black}{\small{}\includegraphics{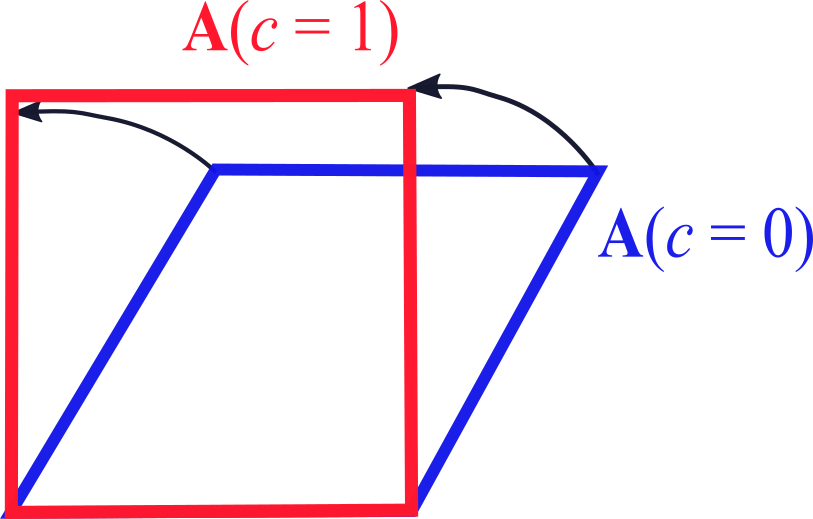}}{\small \par}
\centering{}\textcolor{black}{\small{}}{\small \par}}\textcolor{black}{\small{}\hfill{}}\subfloat[\label{fig:1b}]{\textcolor{black}{\small{}\includegraphics[width=9cm]{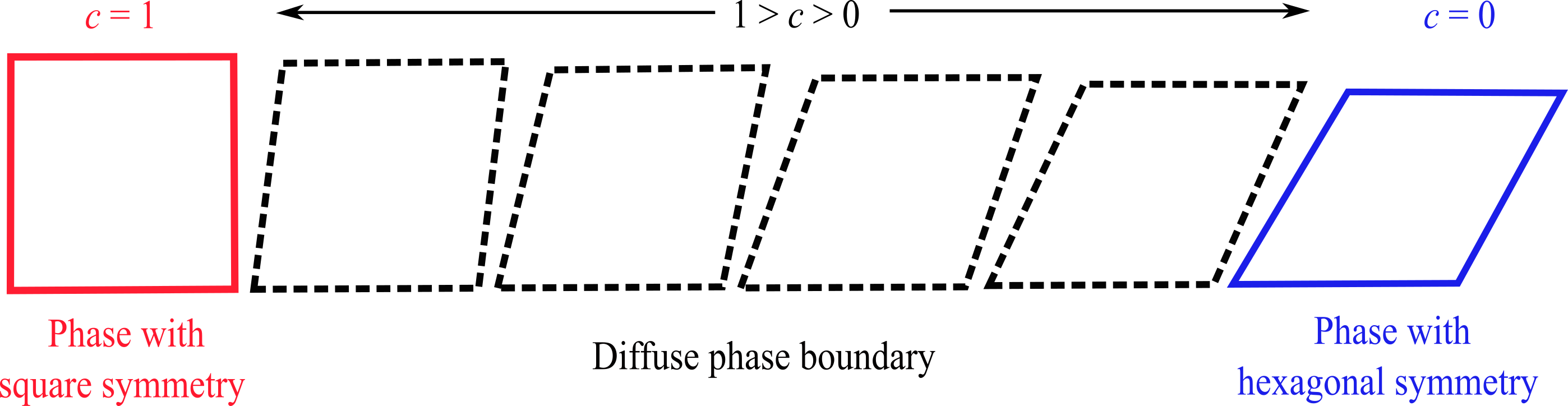}}{\small \par}

\textcolor{black}{\small{}}{\small \par}}\textcolor{black}{\small{}\caption{{\small{}\label{Fig1a-b}Schematic representations of the Cahn-Hilliard
\textendash{} phase field crystal (CH-PFC) method. (a) The reference
hexagonal symmetry (in blue) deforms to a square symmetry (in red)
under the transformation matrix described by Eq. \ref{eq:4}, with
composition field $c=1$. (b) The lattice symmetry transforms from
a square to a hexagonal }\textcolor{black}{\small{}symmetry as a function
of the composition field.}{\small{} The dashed quadrilaterals across
the diffuse phase-boundary illustrate the intermediate lattice symmetries.}}
}
\end{figure}
{\small \par}

\textcolor{black}{\small{}Fig. \ref{Fig1a-b} shows a schematic illustration
of the Cahn-Hilliard \textendash{} phase field crystal concept. In
Fig. \ref{fig:1a}, the transformation matrix describes lattice symmetry
as a function of the composition field. For $c=0$, the transformation
matrix $\mathbf{A}(c=0)$ is an identity matrix, which describes the
composition-dependent Laplacian $\nabla_{c}^{2}$ (in Eq. \ref{eq:3})
in an isotropic coordinate space. With $\mathbf{A}(c=0)$ the CH-PFC
model stabilizes a hexagonal lattice symmetry at equilibrium, see
blue hexagonal symmetry in Fig. \ref{fig:1a}. However, for a system
with $c=1$, the transformation matrix $\mathrm{\mathbf{A}}(c=1)$
introduces anisotropy in the transformation coefficients (in Eq. \ref{eq:3}),
which models the composition-dependent Laplacian in a transformed
coordinate space. With $\mathrm{\mathbf{A}}(c=1)$ the CH-PFC model
results in a square symmetry at equilibrium, see the red square in
Fig. \ref{fig:1a}. Next, Fig. \ref{fig:1b} schematically illustrates
how the CH-PFC model interpolates the lattice symmetry across a diffuse
phase boundary. Here, the transformation matrix is locally defined
in space as a function of the composition field. For $0<c<1$, the
transformation matrix $\mathbf{A}(0<c<1)$ interpolates the peak density
field to describe intermediate lattice symmetries between the square
and the hexagonal, see the dashed quadrilaterals in Fig. \ref{fig:1b}.}{\small \par}

\textcolor{black}{\small{}Next, we describe the evolution of the two
order parameters during phase transition. Here, we assume that the
elastic relaxation of the dimensionless peak density field, $\psi$,
is achieved instantaneously in comparison to the evolution of the
composition field. Consequently, we model $\frac{\delta\mathcal{F}}{\delta\psi}\approx0$
to be maintained throughout the phase transition process. }{\small \par}

\textcolor{black}{\small{}The composition field evolves using a generalized
Cahn-Hilliard equation:}{\small \par}

\textcolor{black}{\small{}
\begin{align}
\frac{\partial c}{\partial\tau} & =-\nabla^{2}\frac{\delta\mathcal{F}}{\delta c}\nonumber \\
 & =-\nabla^{2}(\gamma\frac{\psi}{2}\frac{\partial(\nabla_{c}^{4}+2\nabla_{c}^{2})\psi}{\partial c}+4c^{3}-5c^{2}+2c-\nabla^{2}c).\label{eq:5}
\end{align}
}{\small \par}

\textcolor{black}{\small{}Here, $\gamma=1$ and $\tau$ is the dimensionless
time variable $\tau=t\frac{D}{L^{2}}$. $D$ is the isotropic diffusion
coefficient in Eq. \ref{eq:5} and $L$ is the size of the simulation
grid. The variational derivative in Eq. \ref{eq:5}, produces coupled
terms connecting the peak density field and the composition field.
In Eq. \ref{eq:5}, it is of interest to note the two types of Laplace
operators, $\nabla^{2}\mathrm{\thinspace and\thinspace}\nabla_{c}^{2}$,
respectively. The Laplace operator $\nabla^{2}$ is $\frac{\partial^{2}}{\partial x^{2}}+\frac{\partial^{2}}{\partial y^{2}}$.
This Laplacian computes the Cahn-Hilliard diffusion isotropically.
The composition-dependent Laplacian $\nabla_{c}^{2}$ describes its
partial derivatives in a transformed-coordinate space, see Eq. \ref{eq:3}.
The transformation coefficients are influenced by the local composition
field values and $\nabla_{c}^{2}$ computes the derivatives of $\psi$
in a transformed-coordinate space. The propagation of the composition
diffusion front given by Eq. \ref{eq:5} is affected by both the coarse-grained
lattice arrangements and the local-composition of the model system.
As the composition field evolves, the transformation coefficients
in the composition-dependent Laplacian $\nabla_{c}^{2}$, are updated
accordingly. }{\small \par}

\textcolor{black}{\small{}As the elastic relaxation is much faster
than composition evolution, we introduce a time-like fictive variable
$n$ to compute $\frac{\delta\mathcal{F}}{\delta\psi}\approx0$. The
variable $n$ is treated as a rapidly changing parameter in comparison
to the dimensionless time, $\tau$. This variable $n$ is used as
a relaxation parameter to approximate equilibrium of $\psi$ at each
$c(\tau)$: }{\small \par}

\textcolor{black}{\small{}
\begin{align}
\frac{\partial\psi}{\partial n} & =-\frac{\delta\mathcal{F}}{\delta\psi}+\frac{1}{n_{x}n_{y}}\int\frac{\delta\mathcal{F}}{\delta\psi}d\vec{x}\nonumber \\
 & =\{-\gamma[(r+(1+\nabla_{c}^{2})^{2})\hspace{0.1cm}\psi+\psi^{3}]\nonumber \\
 & \ \ +\frac{1}{n_{x}n_{y}}\int\gamma[(r+(1+\nabla_{c}^{2})^{2})\hspace{0.1cm}\psi+\psi^{3}]\}d\vec{x}.\label{eq:6}
\end{align}
}{\small \par}

\textcolor{black}{\small{}Here, $n_{x}$ and $n_{y}$ are the sides
of a rectangular simulation domain, and $\gamma=1$. Eq. \ref{eq:6}
follows from the numerical scheme introduced by Melenthin et al. \citep{MellenthinJ.KarmaA.andPlapp2008}
that allows equilibrium states to be attained faster in comparison
to the standard equation of motion of the PFC model \citep{Elder2004a}.
Here, $\psi$, is treated as a locally nonconserved order parameter,
while the mass conservation, $\int\psi d\vec{x}$, is ensured globally.
Other approaches to model faster dynamics for the peak density field
can be found in the work by Heinonen et al. \citep{HeinonenV.AchimC.V.KosterlitzJ.M.YingS.C.LowengrubJ.andAla-Nissila2016}.
Note, the variational derivative in Eq. \ref{eq:6} introduces coupled
composition-lattice symmetry terms. These coupled terms affect the
symmetry of the periodic system. }{\small \par}

\textcolor{black}{\small{}Eqs. \ref{eq:5} and \ref{eq:6} are computed
using an Euler discretization scheme in a 2D finite-difference framework.
Simulation grids of size $n_{x}\times n_{y}$ are modeled with periodic
boundary conditions and with grid spacings of $\delta x=\delta y=\frac{4\pi}{q_{0}6\sqrt{3}}$.
At each grid point, the dimensionless composition and peak density
fields are represented in their discrete forms as $c_{ij}$ and $\psi_{ij}$
respectively. The dimensionless composition time derivative in Eq.
\ref{eq:4} is computed at regular time steps of $\Delta\tau$, to
track the evolving composition field. At each time step, $\tau+\Delta\tau$,
the transformation coefficients of the Laplace operator $\nabla_{c}^{2}$
, are updated to correspond with the evolving composition field. Next,
the equilibrium lattice symmetry at time $\tau+\Delta\tau$, is identified
by maintaining $\frac{\delta\mathcal{F}}{\delta\psi}\approx0$. This
general numerical procedure is iterated. In other work, we apply the
CH-PFC method to model Li-ion diffusion in electrode materials \citep{Balakrishna2017}. }{\small \par}

\section*{\textcolor{black}{\small{}CH-PFC simulations}}

\textcolor{black}{\small{}In this section we investigate the nature
of the CH-PFC methods by simulating a few representative examples.
First, we explore how the transformation coefficients stabilize hexagonal
and square symmetries as a function of the composition field. Using
the hexagonal and square symmetries as base cases, we next model a
representative binary alloy with diffuse interfaces. Here, we study
how the model interpolates the peak density field across a diffuse
phase boundary. Finally, we simulate a Cahn-Hilliard type of diffusion
for the composition field and model the accompanying structural changes
to the underlying lattice symmetry during a phase transition.}{\small \par}

\subsection*{\textcolor{black}{\small{}Lattice symmetry}}

\textcolor{black}{\small{}At first, we describe two representative
systems (not necessarily a physical system) with homogeneous composition
fields, $c_{ij}=0$ and $c_{ij}=1,$ respectively. The composition
fields are treated as fixed. Using the composition fields as input,
we compute the peak density fields for the periodic systems. These
representative systems will be generalized subsequently in the following
subsections. Note, the peak density field is rapidly evolving with
reference to the composition field dynamics, and is modeled with a
fictive time in the subsequent computations, see Eq. \ref{eq:6}. }{\small \par}

\textcolor{black}{\small{}Two simulation grids of size $100\times100$
are modeled with periodic boundary conditions. The transformation
matrices at each grid point, for the two representative systems with
$c_{ij}=0$ and $c_{ij}=1$ are computed following Eq. \ref{eq:4}:}{\small \par}

\textcolor{black}{\small{}
\begin{align}
\mathbf{\mathbf{A}_{\mathbf{\mathrm{H}}}=A}(c_{ij}=0) & =\left[\begin{array}{cc}
1 & 0\\
0 & 1
\end{array}\right],\nonumber \\
\mathbf{A}_{\mathbf{\mathrm{S}}}=\mathbf{A}(c_{ij}=1) & =\left[\begin{array}{cc}
1 & -1/\sqrt{3}\\
0 & 2/\sqrt{3}
\end{array}\right].\label{eq:7}
\end{align}
}{\small \par}

\textcolor{black}{\small{}Matrices, $\mathbf{A}_{\mathbf{\mathrm{H}}}$
and $\mathbf{A}_{\mathbf{\mathrm{S}}}$ describe the transformation
coefficients to model the hexagonal and square lattice symmetries
respectively. Note, the determinant of the matrices in Eq. \ref{eq:7}
are $\mathrm{det}(\mathbf{A}_{\mathbf{\mathrm{H}}})=1$ and $\mathrm{det}(\mathbf{A}_{\mathbf{\mathrm{S}}})=1.15$
respectively. The difference in the determinants $\mathrm{det}(\mathbf{A}_{\mathbf{\mathrm{S}}})-\mathrm{det}(\mathbf{A}_{\mathrm{H}})=0.15$,
indicates an area change between the square and hexagonal lattices.
This is because, in the current work we model hexagonal and square
symmetries to assume equal lattice spacing of $\frac{4\pi}{q_{0}\sqrt{3}}$.
Therefore the number density of peaks changes with lattice symmetry.}{\small \par}

\textcolor{black}{\small{}}
\begin{figure}[H]
\begin{centering}
\textcolor{black}{\small{}}\subfloat[\label{fig:2a}]{\textcolor{black}{\small{}\includegraphics[width=7cm]{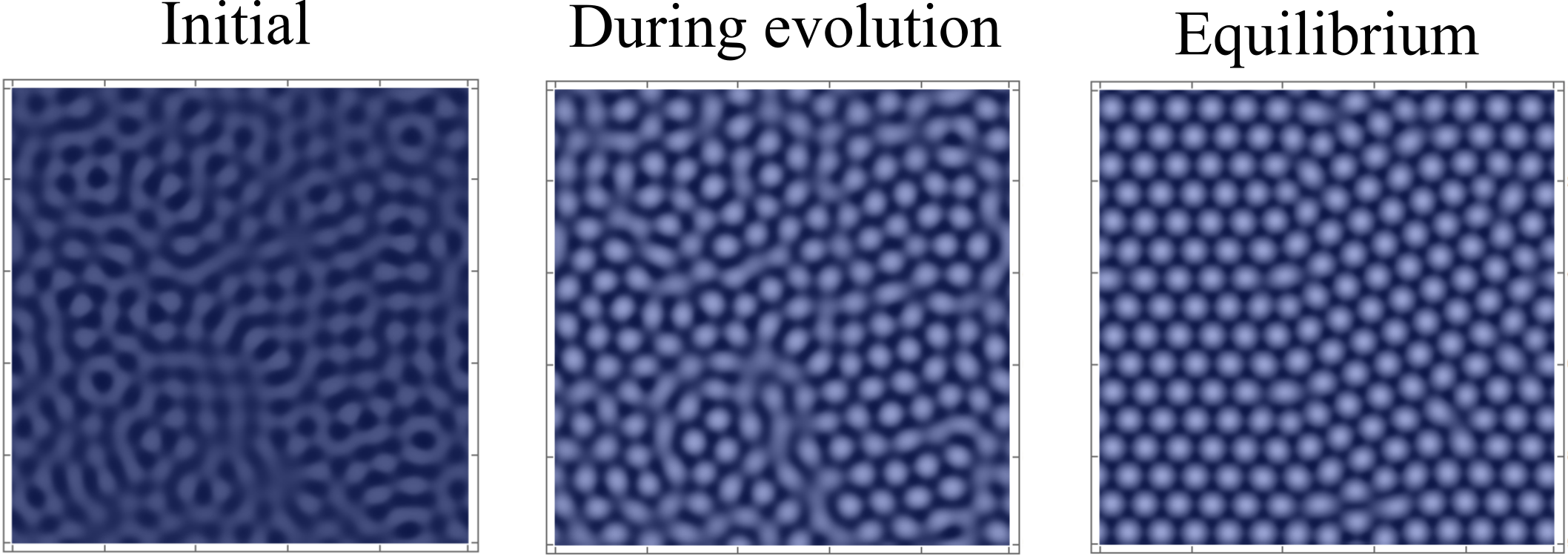}}{\small \par}
\centering{}\textcolor{black}{\small{}}{\small \par}}\textcolor{black}{\small{}\hfill{}}\subfloat[\label{fig:2b}]{\textcolor{black}{\small{}\includegraphics[width=7cm]{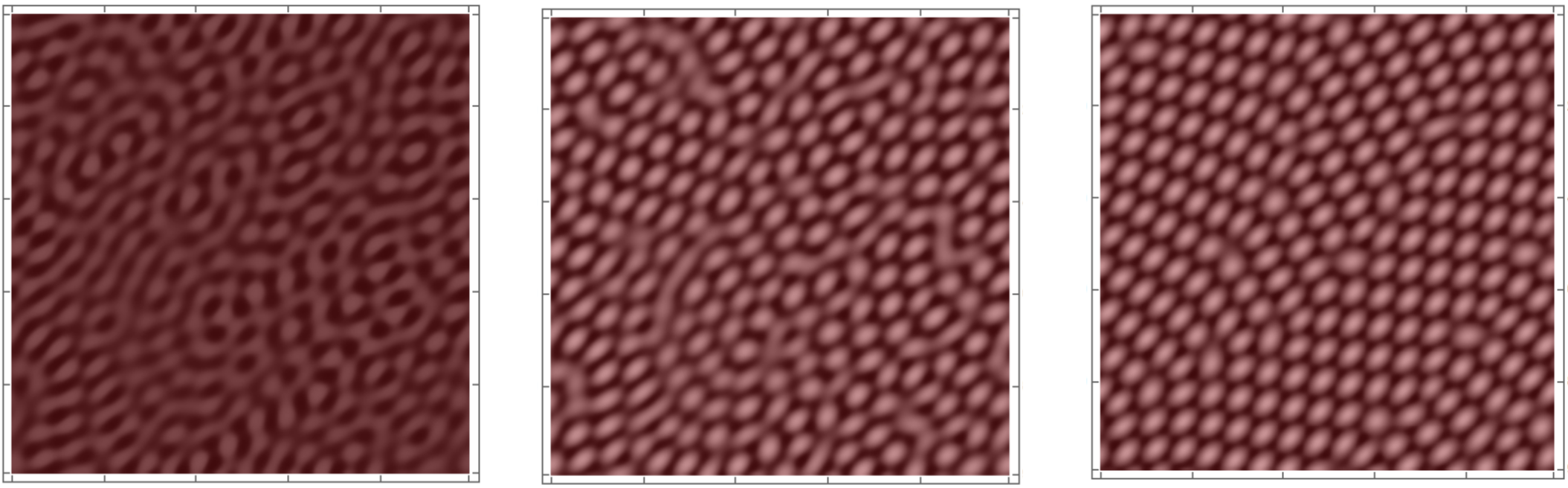}}{\small \par}

\textcolor{black}{\small{}}{\small \par}}\textcolor{black}{\small{}\medskip{}
}
\par\end{centering}{\small \par}
\begin{centering}
\textcolor{black}{\small{}\includegraphics[width=4cm]{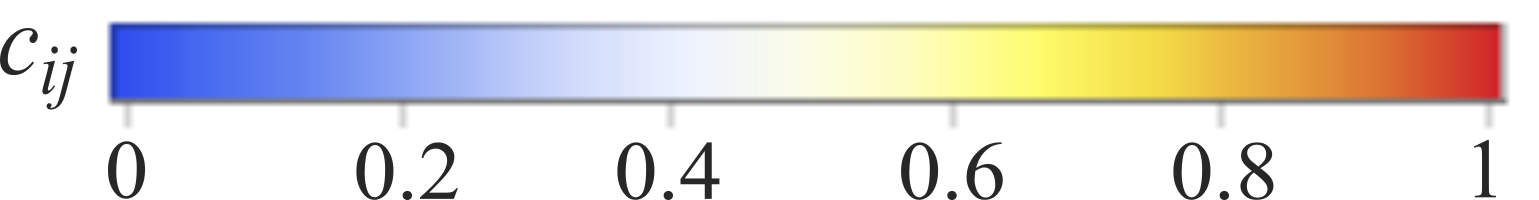}}
\par\end{centering}{\small \par}
\centering{}\textcolor{black}{\small{}\caption{{\small{}\label{fig:2a-b}Evolution of the peak density fields in
representative systems forming (a) hexagonal and (b) square symmetries
with homogeneous}\textcolor{black}{\small{} }{\small{}composition
fields}\textcolor{black}{\small{} $c_{ij}=0$ and $c_{ij}=1$ respectively}{\small{}.
Subfigures illustrate the peak density fields starting from a randomized
initial state (far left), during evolution (centre) and at the final
equilibrium state (far right).}}
}{\small \par}
\end{figure}
{\small \par}

\textcolor{black}{\small{}Using the transformation matrices in Eq.
\ref{eq:7} we next compute the peak density fields of the periodic
systems. The simulation grids are initialized with random peak-density
field values, $-0.1\leq\psi_{ij}\leq0.5$ \textendash{} a condition
that we will refer to as the ``random initial seed''. Starting from
this random state and average density, $\overline{\psi_{ij}}=0.2$,
the evolution of the peak density field, Eq. \ref{eq:6}, is iterated
until equilibrium is reached. }{\small \par}

\textcolor{black}{\small{}Fig. \ref{fig:2a-b} shows the evolution
of density fields from randomized initial states, for the two homogeneous
composition fields, $c_{ij}=0$ and $c_{ij}=1$, respectively. During
evolution, individual grains with hexagonal and square lattice symmetries
nucleate in Fig. \ref{fig:2a} and Fig. \ref{fig:2b} respectively.
Note, grains of different sizes and lattice orientations form during
a CH-PFC simulation, see 'During evolution' in Fig. \ref{fig:2a-b}.
At the grain boundaries, lattice symmetries distort to form coherent
interfaces. At equilibrium, individual grains arrange to minimize
lattice misfits at the grain boundaries. A coarse-grained representation
of hexagonal and square lattice symmetries are formed in Fig. \ref{fig:2a}
and Fig. \ref{fig:2b} respectively. }{\small \par}

\textcolor{black}{\small{}Fig. \ref{fig:2a-b} shows the formation
of multiple grains in homogeneous composition fields and identifies
the position/orientation of the grain boundaries in the model system.
In Fig. \ref{fig:2a}, the  density peaks that model the hexagonal
symmetry are of circular shape. However, for the square symmetry in
Fig. \ref{fig:2b}, the density peaks are ellipsoidal in shape. This
difference in the density peak shapes is explained from the use of
transformation matrices $\mathbf{A_{\mathrm{H}}}$ and $\mathbf{A_{\mathrm{S}}}$
in Eq. \ref{eq:7}. The transformation matrix for hexagonal symmetry,
$\mathbf{A_{\mathrm{H}}}$ describes an isotropic composition-dependent
Laplacian, $\nabla_{c}^{2}$. This computes the density peaks to be
of circular shape. While, the transformation matrix for a square symmetry,
$\mathbf{A_{\mathrm{S}}}$ introduces transformation coefficients
in the composition-dependent Laplace operator, see Eq. \ref{eq:7}
and Eq. \ref{eq:3}. These transformation coefficients shear the density
peaks to an ellipsoidal shape. Similar ellipsoidal density peaks are
observed in the anisotropic PFC simulations \citep{KundinJ.ChoudharyM.A.andEmmerich2014,PrielerR.HubertJ.LiD.VerleyeB.HaberkernR.andEmmerich2009}.
Furthermore, the density peaks near grain boundaries in both Fig.
\ref{fig:2a} and Fig. \ref{fig:2b}, appear smeared and deviate from
the regular ellipsoidal/circular shapes. Here, an interpretation is
that the smeared appearance indicates lattice distortion at the interfaces
to maintain coherency between neighboring grains.}\textcolor{red}{\small{} }{\small \par}

\subsection*{\textcolor{black}{\small{}Diffuse interface}}

\textcolor{black}{\small{}Next, we investigate the model behaviour
to interpolate the peak density field across a diffuse interface in
a representative binary alloy. Here, the hexagonal and square lattice
symmetries at compositions $c_{ij}=0$ and $c_{ij}=1$ are used as
base cases, and correspond to the two phases of the binary alloy.
A representative binary alloy with diffuse phase boundaries is modeled
and its composition field is treated to be fixed. The equilibrium
lattice symmetry for this system with heterogeneous composition field
is computed.}{\small \par}

\textcolor{black}{\small{}A periodic simulation grid of size $200\times30$
is modeled. Here, two phases with $c_{ij}=0$ and $c_{ij}=1$ separated
by a sharp interface is assumed in the initial state:}{\small \par}

\begin{figure}[H]
\begin{centering}
{\small{}\includegraphics[width=7cm]{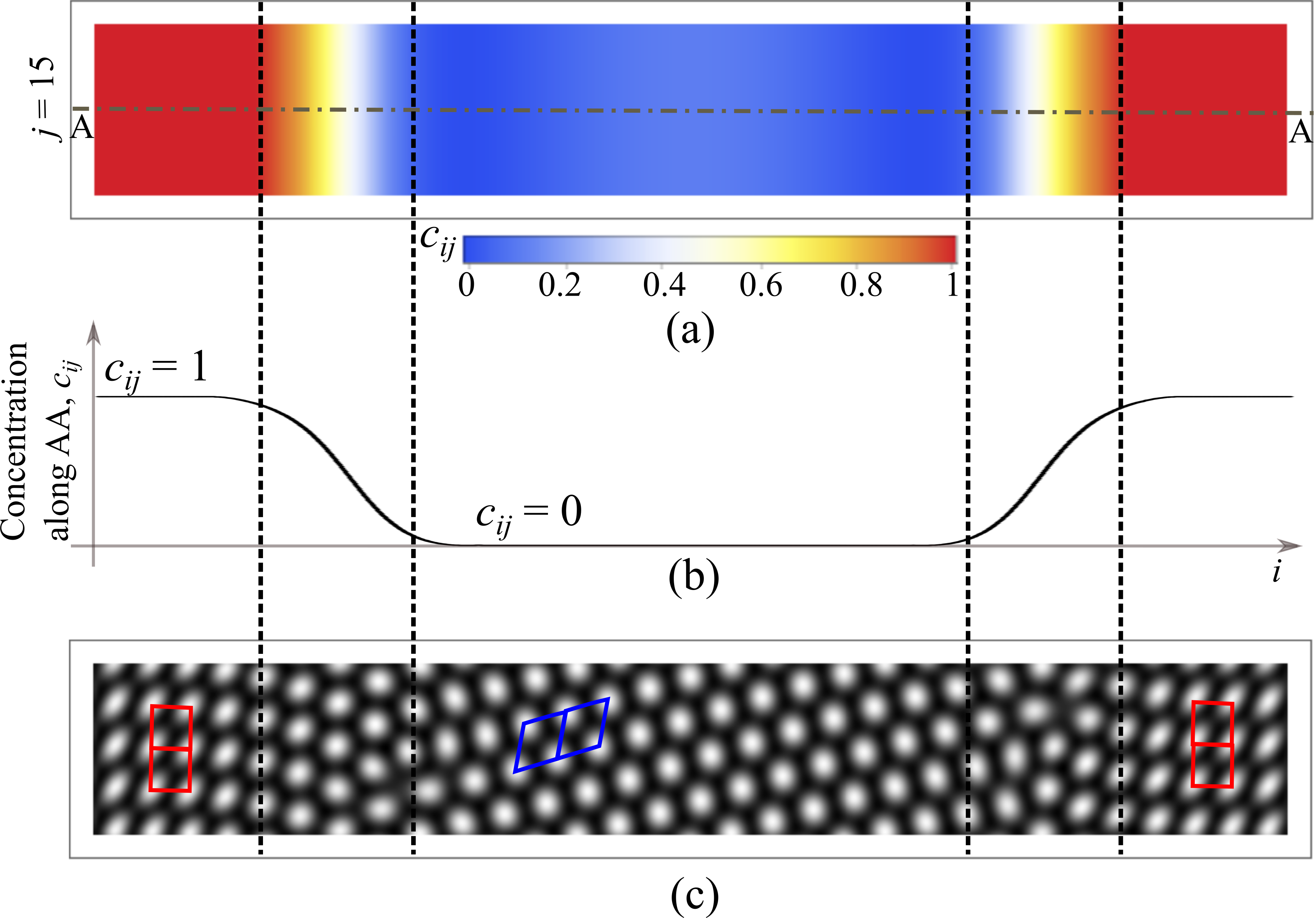}}
\par\end{centering}{\small \par}
\centering{}\textcolor{black}{\small{}\caption{\label{fig:3a-c}{\small{}(a) The composition field of a representative
binary alloy. (b) Variation of composition across AA in the simulation
grid at $j=15$. (c) The equilibrium peak density field describing
the underlying symmetry of the binary alloy corresponding to the composition
field in subfigure 3(a). Square and hexagonal symmetries are described
for phases with $c_{ij}=1$ and $c_{ij}=0$ respectively. The width
of the composition phase boundary (illustrated by vertical dashed-lines)
is numerically calibrated to span across $\sim4$ peaks.}}
}{\small \par}
\end{figure}

\textcolor{black}{\small{}
\begin{equation}
c_{ij}=\left\{ \begin{array}{c}
1\\
0
\end{array}\right.for\begin{array}{c}
i<20,i>180\\
20\leq i\leq180
\end{array}.
\end{equation}
 Next, the composition field is evolved following Eq. \ref{eq:5},
without any influence from the peak density field. That is, $\frac{\partial c}{\partial\tau}=-\nabla^{2}\frac{\delta\mathcal{F}(c,\psi=0)}{\delta c}$.
The composition time derivative is iterated until the phase boundary
begins to smooth and is then held fixed. This is to explore the coupling
of the fast kinetics of $\psi$ for a single interation of $c$. Fig.
\ref{fig:3a-c}(a) illustrates the composition of a  binary alloy
with diffuse phase boundaries. Fig. \ref{fig:3a-c}(b) shows the composition
variation across the simulation grid at $j=15$. }{\small \par}

\textcolor{black}{\small{}}
{\small \par}

\textcolor{black}{\small{}Following Eq. \ref{eq:3}-\ref{eq:4}, the
transformation matrix, $\mathbf{A}(c_{ij})$, is next computed with
$c_{ij}$ describing the discrete composition field shown in Fig.
\ref{fig:3a-c}(a):}{\small \par}

\textcolor{black}{\small{}
\begin{equation}
\mathbf{A\mathrm{(\mathit{c_{ij}\mathrm{)}}}}=\left[{\begin{array}{cc}
1 & \frac{2}{\sqrt{3}}\hspace{0.1cm}\mathrm{cos}[\frac{\pi}{3}+\frac{\pi}{6}c_{ij}]-\frac{1}{\sqrt{3}}\\
0 & \frac{2}{\sqrt{3}}\hspace{0.1cm}\mathrm{sin}[\frac{\pi}{3}+\frac{\pi}{6}c_{ij}]
\end{array}}\right].\label{eq:9}
\end{equation}
}{\small \par}

\textcolor{black}{\small{}Here, $\mathbf{A}(c_{ij})$ defines the
transformed space for the composition-dependent Laplace operator at
each grid point. Using this transformation matrix as an input, the
equilibrium peak density field is next computed, Eq. \ref{eq:6}.}{\small \par}

\textcolor{black}{\small{}To model the lattice symmetry of the binary
alloy shown in Fig. \ref{fig:3a-c}(a), the simulation grid is initialized
with random peak density field values, $-0.1\leq\psi_{ij}\leq0.5$.
Using $\mathbf{A}(c_{ij})$ from Eq. \ref{eq:9}, the evolution of
the peak density field, Eq. \ref{eq:6}, is iterated to find the equilibrium
lattice-symmetry for the model system. Fig. \ref{fig:3a-c}(c) shows
the equilibrium lattice-arrangements described for the heterogeneous
composition field (shown in Fig. \ref{fig:3a-c}(a)). Lattices with
square symmetry are stabilized in the phase with $c_{ij}=1$, and
hexagonal symmetry is observed in the phase with $c_{ij}=0$. At the
phase boundaries, $0<c_{ij}<1$, the coupled CH-PFC model describes
a coarse-grained representation of deformed lattices. Here, the density
peaks are smeared to illustrate the lattice distortion at the phase
boundaries, see Fig. \ref{fig:3a-c}(c). Note, the composition phase
boundary is numerically calibrated to span over $\sim4$ density peaks
(about 25 grid spacings). Fig. \ref{fig:3a-c} provides an atomistic
insight into the coarse-grained lattice arrangements across a diffuse
phase boundary.}{\small \par}

\subsection*{\textcolor{black}{\small{}Phase transition}}

\textcolor{black}{\small{}}
\begin{figure}[H]
\begin{centering}
\textcolor{black}{\small{}}\subfloat[\label{fig:4a}]{\textcolor{black}{\small{}\includegraphics[width=7cm]{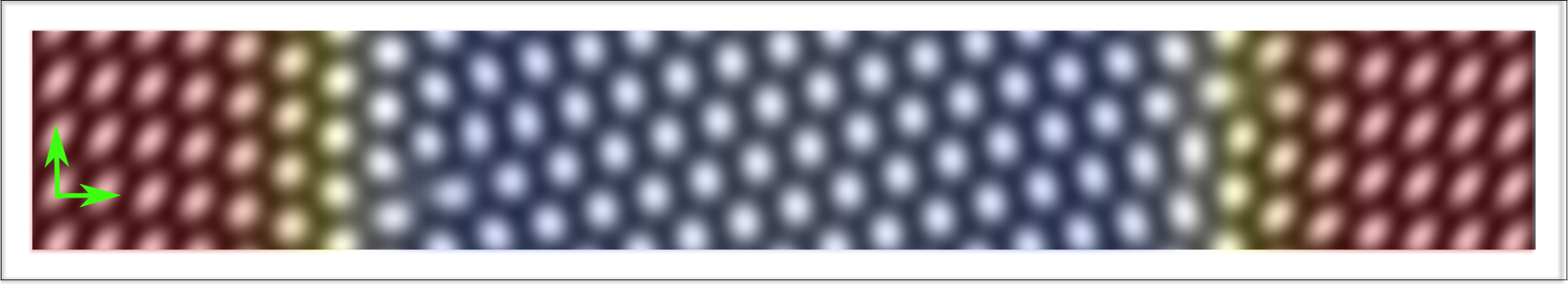}}{\small \par}

\textcolor{black}{\small{}}{\small \par}}\textcolor{black}{\small{}\hfill{}}\subfloat[\label{fig:4b}]{\textcolor{black}{\small{}\includegraphics[width=7cm]{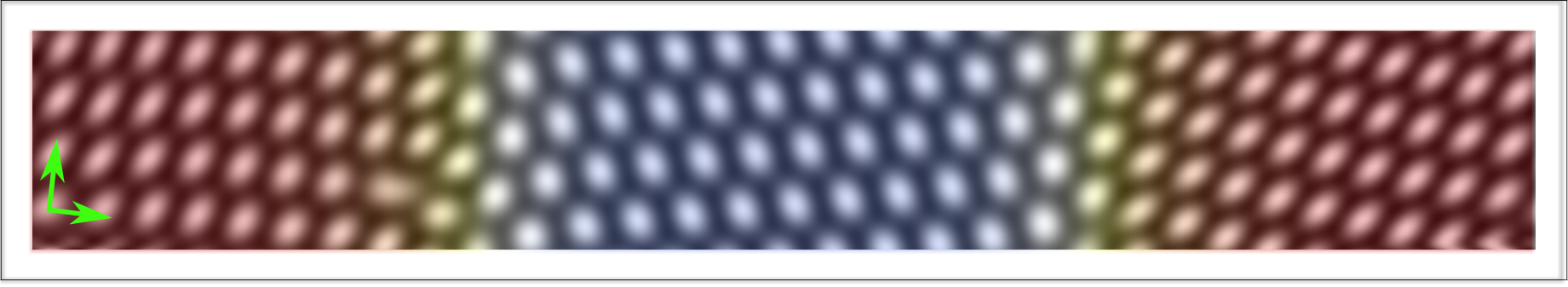}}{\small \par}

\textcolor{black}{\small{}}{\small \par}}
\par\end{centering}{\small \par}
\begin{centering}
\textcolor{black}{\small{}}\subfloat[\label{fig:4c}]{\textcolor{black}{\small{}\includegraphics[width=7cm]{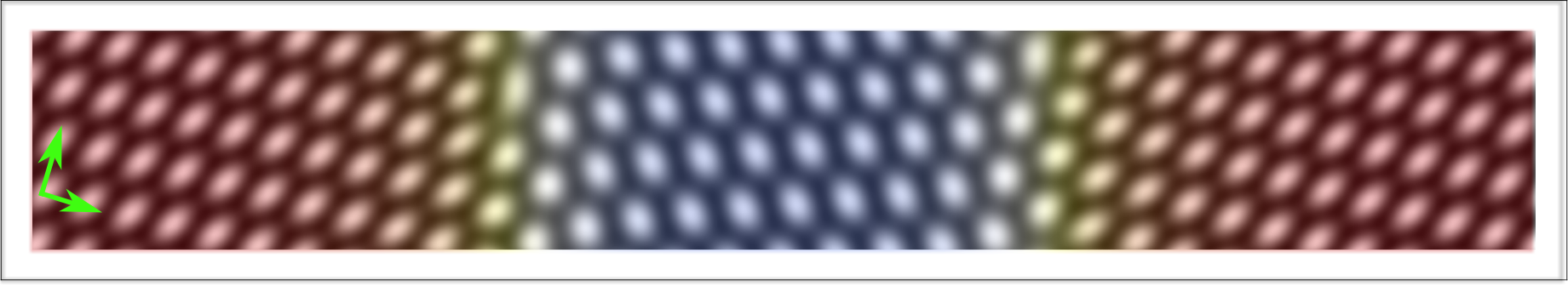}}{\small \par}

\textcolor{black}{\small{}}{\small \par}}\textcolor{black}{\small{}\hfill{}}\subfloat[\label{fig:4d}]{\textcolor{black}{\small{}\includegraphics[width=7cm]{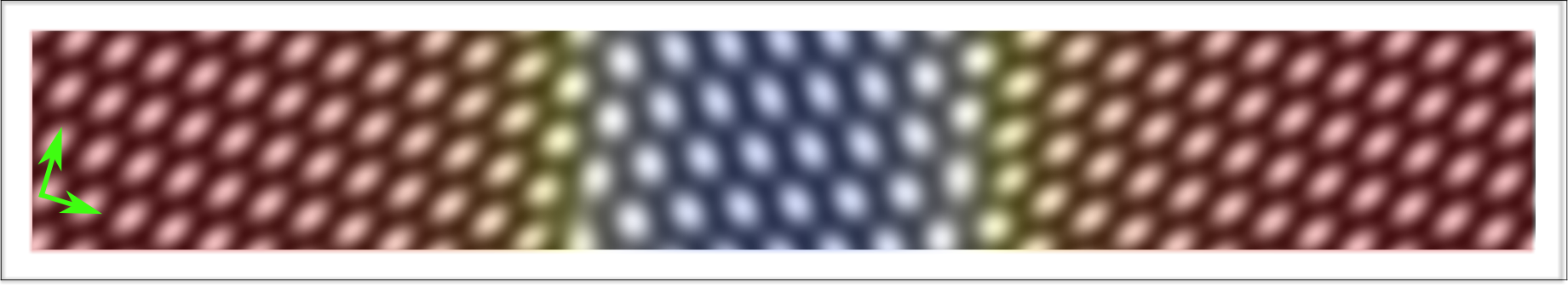}}{\small \par}

\textcolor{black}{\small{}}{\small \par}}
\par\end{centering}{\small \par}
\begin{centering}
\textcolor{black}{\small{}}\subfloat[\label{fig:4e}]{\textcolor{black}{\small{}\includegraphics[width=7cm]{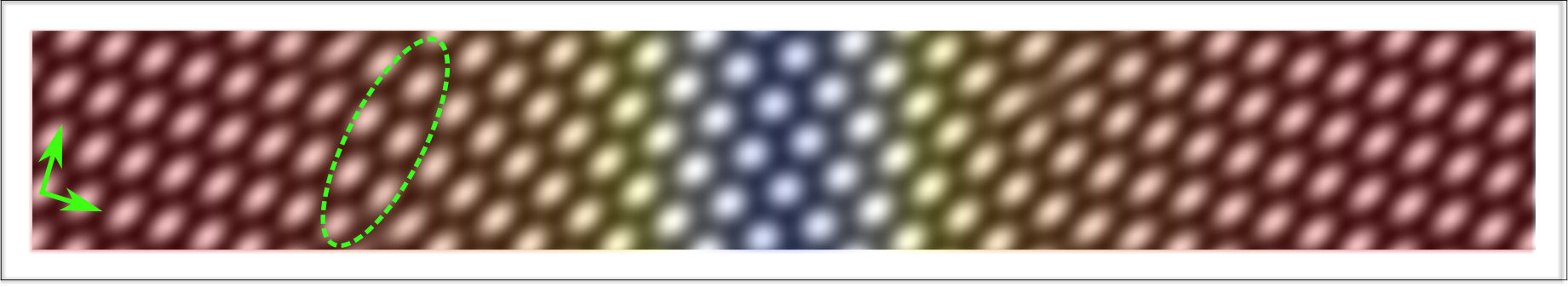}}{\small \par}

\textcolor{black}{\small{}}{\small \par}}\textcolor{black}{\small{}\hfill{}}\subfloat[\label{fig:4f}]{\textcolor{black}{\small{}\includegraphics[width=7cm]{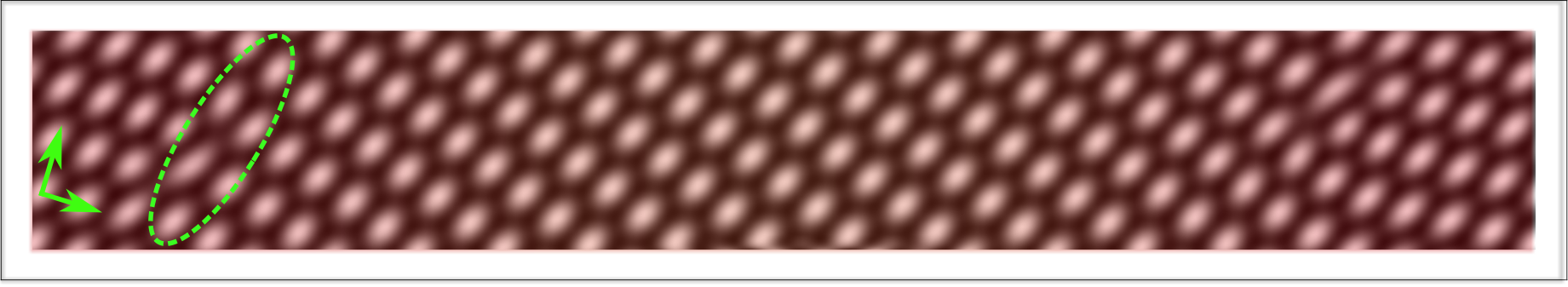}}{\small \par}

\textcolor{black}{\small{}}{\small \par}}
\par\end{centering}{\small \par}
\begin{centering}
\textcolor{black}{\small{}\medskip{}
}
\par\end{centering}{\small \par}
\begin{centering}
\textcolor{black}{\small{}\includegraphics[width=4cm]{4_Users_ananyabalakrishna_Desktop_Master_code_Legend_hor.png}}
\par\end{centering}{\small \par}
\textcolor{black}{\small{}\caption{{\small{}\label{fig:4a-f}Phase transition showing structural evolution
of lattices from a hexagonal (in blue), to a square }\textcolor{black}{\small{}symmetry}{\small{}
(in red). Subfigures illustrate lattice transformations as a function
of the dimensionless composition field, $c_{ij}$. The green arrows
indicate the orientation of the square symmetry during phase transition.
The dashed line in subfigures 4(e-f) indicate a grain boundary in
the square symmetry phase.}}
}{\small \par}
\end{figure}
{\small \par}

\textcolor{black}{\small{}Up to this point, we only modeled the microscopic
configurations at fixed compositions. However, to model phase transition
with microscopic insights on the coarse-grained lattice symmetry,
we need to simulate the evolution of the composition field. The binary
alloy in Fig. \ref{fig:3a-c} is considered as the initial state,
and we next extend the simulation to describe the propagation of the
diffusion front. A representative Cahn-Hilliard type of diffusion
for the composition field is modeled. During the phase transition,
the equilibrium lattice arrangements of the underlying system is computed.
An assumption made in this simulation is that the dynamics of elastic
relaxation (equilibrating the peak density field) is several times
faster than the diffusion of the composition field. Using this CH-PFC
approach we investigate how composition field influences the lattice
arrangements in a model system during phase transitions.}{\small \par}

{\small{}Taking as an initial state, the lattice arrangements described
for a binary alloy from Fig. \ref{fig:3a-c}, the phase transition
is modeled by allowing composition to diffuse into the simulation
domain. }\textcolor{black}{\small{}The composition field is held fixed
at $c_{ij}=1$, for $i<20$ and $i>180$ throughout the simulation.
This boundary condition is a proxy for having a consistent composition
reservoir. The composition field on the remaining part of the simulation
grid, $20\leq i\leq180$, is allowed to evolve with time. The composition
time derivative, Eq. \ref{eq:5}, is iterated from $\tau=0$ to $\tau=2500$,
in dimensionless time intervals of $\Delta\tau=25$. Note, the composition
evolution at $\tau$, receives input from the equilibrium peak density
field calculated for the $(\tau-1)$ time step. The composition field
is tracked as $c_{ij}(\tau+\Delta\tau)=c_{ij}(\tau)+\Delta\tau\frac{\partial c_{ij}}{\partial\tau}$,
until a homogenous phase is obtained.}{\small \par}

\textcolor{black}{\small{}The composition field at each evolution
step, $c_{ij}(\tau)$, is used as an input to compute the transformation
matrix in Eq. \ref{eq:9}. At a given time step, $\tau$, the transformation
matrix $\mathbf{A}(c_{ij}(\tau))$, is used to calculate the equilibrium
peak-density-field following Eq. \ref{eq:6}. The composition and
peak density fields are iterated until the phase transition is complete.}{\small \par}

\textcolor{black}{\small{}Fig. \ref{fig:4a-f} shows the structural
evolution of the coarse-grained lattice arrangements during the phase
transition. At the intial state $\tau=0$, the coarse-grained lattice
symmetry for the heterogeneous composition, $c_{ij}(\tau=0)$ is described,
see Fig. \ref{fig:4a}. Here, two coherent phases with square and
hexagonal symmetries are formed in domains with $c_{ij}=1$ and $c_{ij}=0$
respectively. Note in Fig. \ref{fig:4a}, the edges of the square
lattices are mostly aligned with the axes of the simulation grid.
A pair of green arrows in Fig. \ref{fig:4a} illustrates the orientation
of square lattices in the simulation grid. Across the diffuse phase
boundary, hexagonal and square lattices are distorted to maintain
coherency, see Fig. \ref{fig:4a}. Next, in Fig. \ref{fig:4a-f}(b-e),
as the composition field diffuses into the simulation domain, the
hexagonal lattice symmetry is transformed to a square symmetry. }{\small \par}

\textcolor{black}{\small{}In Fig. \ref{fig:4b}, the phase with square
symmetry occupies $\sim50\%$ of the simulation grid. Here, it is
interesting to note that square lattices begin to rotate uniformly
as the diffusion front propagates through the simulation grid. In
Fig. \ref{fig:4a-f}(c-d), the square lattice symmetry is observed
to rotate further (e.g., orientation of the green arrows in Fig. \ref{fig:4a-f}(c-d)).
We interpret that the square lattices rotate to maintain coherency
with the neighboring hexagonal phase. Note, the periodic boundary
conditions on the simulation domain further enforce an additional
strain on the peak density field. This is discussed in detail in the
next section of this paper. In Fig. \ref{fig:4e}, a grain boundary
(as indicated by the dashed line) is formed in the square symmetry
phase. This grain boundary migrates in the square symmetry phase and
remains in the homogeneous phase, see Fig. \ref{fig:4f}. At $\tau=2500$,
the phase transition is complete with a homogenous composition field
and a phase with square symmetry is described at equilibrium, see
Fig. \ref{fig:4f}.}{\small \par}

\section*{\textcolor{black}{\small{}Discussion of the CH-PFC model}}

\textcolor{black}{\small{}The coupled Cahn-Hilliard \textendash{}
phase field crystal model provides a theoretical framework to describe
continuum phase transition with microscopic insights. There are several
issues we feel remain to be clarified in interpreting the simulations.
Among these issues are three questions: Do the peaks in the CH-PFC
simulations represent atomic sites or illustrate the underlying lattice
symmetry? Are the total number of peaks in a simulation grid conserved?
In Eq. \ref{eq:1} why was the composition field coupled with the
peak density field only via the Laplace operator? In this section,
we discuss these key details of the coupled CH-PFC model and explore
potential further work.}{\small \par}

\begin{figure}[H]
\begin{centering}
{\small{}\includegraphics[width=7cm]{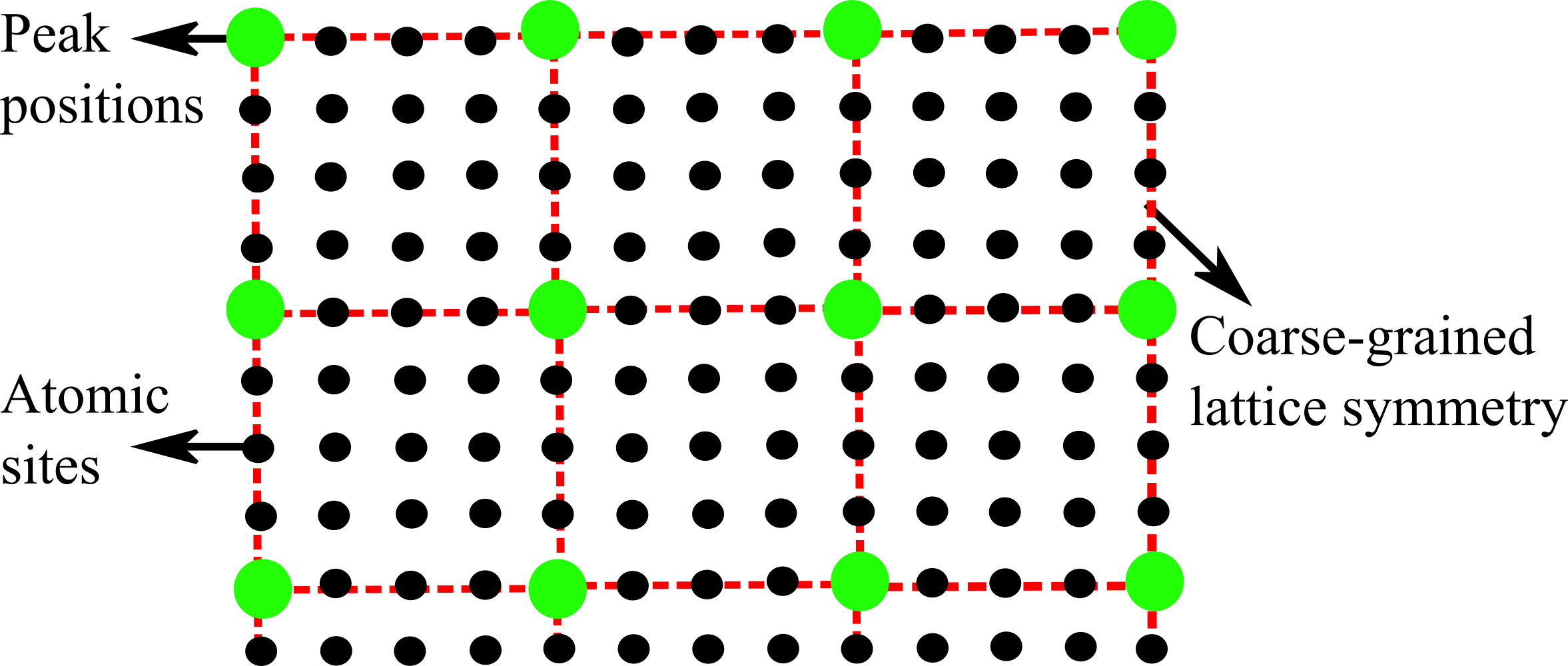}}
\par\end{centering}{\small \par}
\begin{centering}
\textcolor{black}{\small{}\smallskip{}
}
\par\end{centering}{\small \par}
{\small{}\caption{\label{fig:5}{\small{}Schematic illustration of the atomic sites,
peak positions and coarse-grained lattice symmetry in a square phase
system. The small-black dots represent atomic sites of the unit cell.
The big-green dots schematically indicate peak positions in the CH-PFC
simulation. The dashed-red lines connecting the peaks highlight an
example of a coarse-grained lattice symmetry of the underlying unit
cells. Note, in this figure the coarse grained symmetry is four times
the unit cell size. However, in our CH-PFC simulations the coarse-grained
symmetry is multiple times larger than a unit cell.}}
}{\small \par}

\end{figure}

{\small{}First, the peak density field}\textcolor{black}{\small{}
in CH-PFC simulation}{\small{} describes the coarse-grained lattice
symmetry of the underlying atomic arrangements. Individual peaks do
not represent atomic sites, however the arrangement of peaks indicates
the unit cell symmetry of the model system. }\textcolor{black}{\small{}Similarly,
a grain boundary in a CH-PFC simulation is a coarse-grained approximation
of the underlying lattice orientations, distortions and defects.}{\small{}
Fig. \ref{fig:5} provides a schematic illustration of the difference
between atomic sites, peak positions and coarse-grained lattice symmetry.
In Fig. \ref{fig:5}, the small-black dots indicate atomic sites,
which correspond to the deterministic positions of atoms in the unit
cell. The big-green dots highlight representative peak positions modeled
by a CH-PFC method. The dashed-red lines connecting the peaks in Fig.
\ref{fig:5}, indicate an example of a coarse-grained lattice symmetry.
In Fig. \ref{fig:5}, the side of the coarse-grained lattice is four
times that of the unit cell. However, in our CH-PFC simulations, the
coarse-grained lattice is several times larger than a unit cell. }{\small \par}

{\small{}}
{\small \par}

{\small{}Second, the number density of peaks in the CH-PFC simulations
are not necessarily conserved.}\textcolor{black}{\small{} Let us consider
Fig. \ref{fig:2a-b}, where hexagonal and square symmetries are described
on identitical computational grids of size $100\times100$. The total
number of peaks in both these symmetry systems are not necessarily
the same. This can be explained from two reasons: First, the transformation
matrices in Eq. \ref{eq:7}, $\mathbf{A_{\mathrm{H}}}$ and $\mathbf{A_{\mathrm{S}}}$,
describe lattice symmetries with an area difference $(\sim0.15)$.
Second, the periodic boundary conditions on the computational grid
enforces a strain on the peak density field. That is, on an infinitely
large grid size, the peak density field would assume the fundamental
length scale of $\frac{4\pi}{q_{0}\sqrt{3}}$ specified by the transformation
coefficients in Eq. \ref{eq:3}. However, by modeling this density
field on a periodic grid with finite dimensions, we strain the lattice
symmetry spacing and force the peak density field to satisfy periodic
boundary conditions. To minimize these imposed strains and to simultaneously
maintain periodicity, the CH-PFC model introduces (or removes) peaks
to (or from) the simulation grid. For future applications of the CH-PFC
model, the computational grid size is to be calibrated to correspond
to the closest fundamental length scale of the peak density field.
Alternatively, numerical correction terms to Eq. \ref{eq:6} to conserve
the number of peaks can be used \citep{AlandS.HatzikirouH.LowengrubJ.andVoigt2015}.}{\small \par}

\textcolor{black}{\small{}Finally, in Eq. (1), the composition and
the peak density fields are coupled only via the Laplace operator.
That is, the coarse-grained lattice symmetry described by Eq. \ref{eq:1}
is solely determined by the coordinate transformation coefficients
of the Laplace operator. These transformation coefficients (which
are functions of the composition field) describe lattice transformations
with hexagonal symmetry as the reference structure. In this paper,
we assumed the ideal free energy contribution from other non-linear
terms $(\psi^{2},\psi^{4})$ in Eq. \ref{eq:1}, to be independent
of the composition field for a couple of reasons: First, this assumption
allows the CH-PFC model to stabilize a reference hexagonal lattice
symmetry for composition field, $c_{ij}=0$. Second, Eq. \ref{eq:1}
will always describe a crystalline/ordered state for the model system.
This is because the driving force for the peak density field towards
the disordered state (controlled by term $\psi^{4}$) is not a function
of the composition field.}\textit{\textcolor{black}{\small{} }}{\small \par}

\section*{\textcolor{black}{\small{}Summary}}

\textcolor{black}{\small{}We introduced a 2D theoretical framework,
which combined a Cahn-Hilliard (CH) model and a phase field crystal
(PFC) model, to describe a phase transition process. In this CH-PFC
method, the composition field was coupled to the coarse-grained lattice
symmetry (peak density field) of the periodic system. The CH-PFC modeling
approach captured the effects of microscopic configurations, such
as lattice orientations, distortions and presence of defects, on the
phase-transition process. Furthermore, the model described the structural
evolution of the coarse-grained lattice symmetry during a phase change.}{\small \par}

\textcolor{black}{\small{}Using the CH-PFC approach, we stabilized
representative lattice symmetries (hexagonal and square) as a function
of the composition field. Here, we found that multiple grains formed
in a single phase, and identified the position and orientation of
grain boundaries. Next, in a binary alloy, we described the coarse-grained
distortion of lattice symmetry across a diffuse phase boundary. Finally,
we modelled a representative phase transition process \textendash{}
here, the CH-PFC simulations modeled grain rotations and grain boundary
migrations during phase change.}{\small \par}

\section*{\textcolor{black}{\small{}Acknowledgements}}

\textcolor{black}{\small{}A.R.B acknowledges the support of the Lindemann
trust fellowship. The authors would like to acknowledge the support
by the grant DE-SC0002633 funded by the U.S. Department of Energy,
Office of Science, in carrying out this work. Further, the authors
wish to thank Dr. Rachel Zucker for useful discussions on phase field
crystal modeling methods. }{\footnotesize{}}\\
{\footnotesize \par}

{\footnotesize{}\bibliographystyle{unsrt}
\bibliography{13_Users_ananyabalakrishna_Desktop_CHPFC_references}

\begin{thebibliography}{10}

\bibitem{LinesM.E.1977}
Glass A.~M. {Lines M.E.}
\newblock {\em {Principles and applications of ferroelectrics and related
  materials}}.
\newblock Oxford university press, 1977.

\bibitem{James1986}
R.D. James.
\newblock {Displacive phase transformations in solids}.
\newblock {\em J. Mech. Phys. Solids}, 34:359--394, 1986.

\bibitem{Meethong2008}
N.~Meethong, Y.~H. Kao, M.~Tang, H.~Y. Huang, W.~C. Carter, and Y.-M. Chiang.
\newblock {Electrochemically Induced Phase Transformation in Nanoscale Olivines
  Li1-xMPO4 (M = Fe, Mn)}.
\newblock {\em Chem. Mater.}, 20:6189--6198, 2008.

\bibitem{LevanyukA.P.&Sigov1988}
A.~S. {Levanyuk, A. P., {\&} Sigov}.
\newblock {\em {Defects and structural phase transitions}}.
\newblock Gordon and Breach Science Publishers, 1988.

\bibitem{SongY.ChenX.DabadeV.ShieldT.W.&James2013}
R.~D. {Song, Y., Chen, X., Dabade, V., Shield, T. W., {\&} James}.
\newblock {Enhanced reversibility and unusual microstructure of a
  phase-transforming material}.
\newblock {\em Nature}, 502(7469):85--88, 2013.

\bibitem{Nie2014}
A.~Nie, L.-Y. Gan, Y.~Cheng, Q.~Li, Y.~Yuan, F.~Mashayek, H.~Wang, R.~Klie,
  U.~Schwingenschlogl, and R.~Shahbazian-Yassar.
\newblock {Twin boundary-assisted lithium ion transport}.
\newblock {\em Nano Lett.}, 15:610--615, 2014.

\bibitem{YuanY.NieA.OdegardG.M.XuR.ZhouD.SanthanagopalanS.HeK.Asayesh-ArdakaniH.MengD.D.KlieR.F.andJohnson}
Y.~Yuan, A.~Nie, G.M. Odegard, R.~Xu, D.~Zhou, S.~Santhanagopalan, K.~He,
  H.~Asayesh-Ardakani, D.D. Meng, R.F. Klie, and C.~Johnson.
\newblock {Asynchronous crystal cell expansion during lithiation of
  K+-stabilized $\alpha$-MnO2}.
\newblock {\em Nano Lett.}, 15:2998--3007, 2015.

\bibitem{Tang2006a}
M.~Tang, W.C. Carter, and R.M. Cannon.
\newblock {Diffuse interface model for structural transitions of grain
  boundaries}.
\newblock {\em Phys. Rev. B}, 73:024102, 2006.

\bibitem{Warren2003a}
J.A. Warren, R.~Kobayashi, A.E. Lobkovsky, and W.C. Carter.
\newblock {Extending phase field models of solidification to polycrystalline
  materials}.
\newblock {\em Acta Mater.}, 51:6035--6058, 2003.

\bibitem{Tang2009}
M.~Tang, H.Y. Huang, N.~Meethong, Y.H. Kao, W.C. Carter, and Y.M. Chiang.
\newblock {Model for the particle size, overpotential, and strain dependence of
  phase transition pathways in storage electrodes: application to nanoscale
  olivines}.
\newblock {\em Chem. Mater.}, 21:1557--1571, 2009.

\bibitem{Cogswell2012a}
D.A. Cogswell and M.Z. Bazant.
\newblock {Coherency strain and the kinetics of phase separation in LiFePO4
  nanoparticles}.
\newblock {\em ACS Nano}, 6:2215--2225, 2012.

\bibitem{Chen2002}
L.Q. Chen.
\newblock {Phase-field models for microstructure evolution}.
\newblock {\em Annu. Rev. Mater. Res.}, 32:113--140, 2002.

\bibitem{Balakrishna2016}
A.R. Balakrishna, J.E. Huber, and I.~M{\"{u}}nch.
\newblock {Nanoscale periodic domain patterns in tetragonal ferroelectrics: A
  phase-field study}.
\newblock {\em Phys. Rev. B}, 93:174120, 2016.

\bibitem{Elder2002a}
K.R. Elder, M.~Katakowski, M.~Haataja, and M.~Grant.
\newblock {Modeling elasticity in crystal growth}.
\newblock {\em Phys. Rev. Lett.}, 88:245701, 2002.

\bibitem{Elder2004a}
K.R. Elder and M.~Grant.
\newblock {Modeling elastic and plastic deformations in nonequilibrium
  processing using phase field crystals}.
\newblock {\em Phys. Rev. E}, 70:051605, 2004.

\bibitem{Emmerich2012b}
H.~Emmerich, H.~L{\"{o}}wen, R.~Wittkowski, T.~Gruhn, G.I. T{\'{o}}th,
  G.~Tegze, and L.~Gr{\'{a}}n{\'{a}}sy.
\newblock {Phase-field-crystal models for condensed matter dynamics on atomic
  length and diffusive time scales: an overview}.
\newblock {\em Adv. Phys.}, 61:665--743, 2012.

\bibitem{Provatas2007}
N.~Provatas, J.A. Dantzig, B.~Athreya, P.~Chan, P.~Stefanovic, N.~Goldenfeld,
  and K.R. Elder.
\newblock {Using the phase-field crystal method in the multi-scale modeling of
  microstructure evolution.}
\newblock {\em JOM J. Miner. Met. Mater. Soc.}, 59:83--90, 2007.

\bibitem{Tupper2008}
P.F. Tupper and M.~Grant.
\newblock {Phase field crystals as a coarse-graining in time of molecular
  dynamics.}
\newblock {\em EPL (Europhysics Lett.}, 81:40007, 2008.

\bibitem{Seymour2016b}
M.~Seymour and N.~Provatas.
\newblock {Structural phase field crystal approach for modeling graphene and
  other two-dimensional structures.}
\newblock {\em Phys. Rev. B}, 93:035447, 2016.

\bibitem{GranasyL.TegzeG.TothG.I.andPusztai2011}
T.~{Gr{\'{a}}n{\'{a}}sy, L., Tegze, G., T{\'{o}}th, G.I. and Pusztai}.
\newblock {Phase-field crystal modelling of crystal nucleation, heteroepitaxy
  and patterning.,}.
\newblock {\em Philos. Mag.}, 91:123--149, 2011.

\bibitem{ElderK.R.ProvatasN.BerryJ.StefanovicP.andGrant2007a}
M.~{Elder, K.R., Provatas, N., Berry, J., Stefanovic, P. and Grant}.
\newblock {Phase-field crystal modeling and classical density functional theory
  of freezing}.
\newblock {\em Phys. Rev. B}, 75:064107, 2007.

\bibitem{ElderK.R.HuangZ.F.andProvatas2010}
N.~{Elder, K.R., Huang, Z.F. and Provatas}.
\newblock {Amplitude expansion of the binary phase-field-crystal model}.
\newblock {\em Phys. Rev. E}, 81:011602, 2010.

\bibitem{KundinJ.ChoudharyM.A.andEmmerich2014}
H.~{Kundin, J., Choudhary, M.A. and Emmerich}.
\newblock {Bridging the phase-field and phase-field crystal approaches for
  anisotropic material systems}.
\newblock {\em Eur. Phys. J. Spec. Top.}, 223:363--372, 2014.

\bibitem{AlsterE.ElderK.R.HoytJ.J.andVoorhees2017}
P.W. {Alster, E., Elder, K.R., Hoyt, J.J. and Voorhees}.
\newblock {Phase-field-crystal model for ordered crystals}.
\newblock {\em Phys. Rev. E}, 95:022105, 2017.

\bibitem{HeinonenV.AchimC.V.ElderK.R.BuyukdagliS.andAla-Nissila2014}
T.~{Heinonen, V., Achim, C.V., Elder, K.R., Buyukdagli, S. and Ala-Nissila}.
\newblock {Phase-field-crystal models and mechanical equilibrium}.
\newblock {\em Phys. Rev. E}, 89:032411, 2014.

\bibitem{ElderK.R.ProvatasN.BerryJ.StefanovicP.andGrant2007}
M.~{Elder, K.R., Provatas, N., Berry, J., Stefanovic, P. and Grant}.
\newblock {Phase-field crystal modeling and classical density functional theory
  of freezing.}
\newblock {\em Phys. Rev. B}, 75:064107, 2007.

\bibitem{ZhangW.andMi2016}
J.~{Zhang, W. and Mi}.
\newblock {Phase field crystal modelling of the order-to-disordered atomistic
  structure transition of metallic glasses}.
\newblock {\em IOP Conf. Ser. Mater. Sci. Eng.}, 117:012056, 2016.

\bibitem{TaoY.ZhengC.JingZ.Wei-PingD.andLin2012}
W.~{Tao, Y., Zheng, C., Jing, Z., Wei-Ping, D. and Lin}.
\newblock {Effect of grain boundary on spinodal decomposition using the phase
  field crystal method}.
\newblock {\em Chinese Phys. Lett.}, 29:078103, 2012.

\bibitem{GreenwoodM.SinclairC.andMilitzer2012}
M.~{Greenwood, M., Sinclair, C. and Militzer}.
\newblock {Phase field crystal model of solute drag}.
\newblock {\em Acta Mater.}, 60:5752--5761, 2012.

\bibitem{BallJ.M.andJames1987}
R.D. {Ball, J.M. and James}.
\newblock {Fine phase mixtures as minimizers of energy}.
\newblock {\em Arch. Ration. Mech. Anal.}, 100:13--52, 1987.

\bibitem{Bhattacharya2003}
K.~Bhattacharya.
\newblock {\em {Microstructure of martensite: why it forms and how it gives
  rise to the shape-memory effect}}.
\newblock Oxford university press, 2003.

\bibitem{MellenthinJ.KarmaA.andPlapp2008}
M.~{Mellenthin, J., Karma, A. and Plapp}.
\newblock {Phase-field crystal study of grain-boundary premelting}.
\newblock {\em Phys. Rev. B}, 78:184110, 2008.

\bibitem{HeinonenV.AchimC.V.KosterlitzJ.M.YingS.C.LowengrubJ.andAla-Nissila2016}
T.~{Heinonen, V., Achim, C.V., Kosterlitz, J.M., Ying, S.C., Lowengrub, J. and
  Ala-Nissila}.
\newblock {Consistent hydrodynamics for phase field crystals}.
\newblock {\em Phys. Rev. Lett.}, 116:024303, 2016.

\bibitem{Balakrishna2017}
A.R. Balakrishna and W.C. Carter.
\newblock {Modeling phase transition in nanoscale electrodes using the coupled
  Cahn-Hilliard - phase field crystal methods}.
\newblock {\em Prep.}, 2017.

\bibitem{PrielerR.HubertJ.LiD.VerleyeB.HaberkernR.andEmmerich2009}
H.~{Prieler, R., Hubert, J., Li, D., Verleye, B., Haberkern, R. and Emmerich}.
\newblock {An anisotropic phase-field crystal model for heterogeneous
  nucleation of ellipsoidal colloids}.
\newblock {\em J. Phys. Condens. Matter}, 21:464110, 2009.

\bibitem{AlandS.HatzikirouH.LowengrubJ.andVoigt2015}
A.~{Aland, S., Hatzikirou, H., Lowengrub, J. and Voigt}.
\newblock {A mechanistic collective cell model for epithelial colony growth and
  contact inhibition}.
\newblock {\em Biophys. J.}, 109:1347--1357, 2015.

\end{thebibliography}
}{\footnotesize \par}
\end{document}